\documentclass[prl,twocolumn,a4paper,superscriptaddress,floatfix]{revtex4}
\usepackage{graphicx}
\begin{document}

\newcommand{\be}{\begin{equation}}
\newcommand{\ee}{\end{equation}}
\newcommand{\bq}{\begin{eqnarray}}
\newcommand{\eq}{\end{eqnarray}}
\newcommand{\bsq}{\begin{subequations}}
\newcommand{\esq}{\end{subequations}}
\newcommand{\bc}{\begin{center}}
\newcommand{\ec}{\end{center}}
\newcommand {\R}{{\mathcal R}}
\newcommand{\al}{\alpha}
\newcommand\lsim{\mathrel{\rlap{\lower4pt\hbox{\hskip1pt$\sim$}}
    \raise1pt\hbox{$<$}}}
\newcommand\gsim{\mathrel{\rlap{\lower4pt\hbox{\hskip1pt$\sim$}}
    \raise1pt\hbox{$>$}}}

\title{Scaling of cosmological domain wall networks with junctions}

\author{P.P. Avelino} 
\email[Electronic address: ]{ppavelin@fc.up.pt} 
\affiliation{Centro de F\'{\i}sica do Porto, Rua do Campo Alegre 687, 4169-007 Porto, Portugal} 
\affiliation{Departamento de F\'{\i}sica da Faculdade de Ci\^encias 
da Universidade do Porto, Rua do Campo Alegre 687, 4169-007 Porto, Portugal} 
\affiliation{JILA, University of Colorado, Boulder, CO 80309-0440}
\author{C.J.A.P. Martins} 
\email[Electronic address: ]{C.J.A.P.Martins@damtp.cam.ac.uk} 
\affiliation{Centro de F\'{\i}sica do Porto, Rua do Campo Alegre 687, 4169-007 Porto, Portugal} 
\affiliation{Department of Applied Mathematics and Theoretical  
Physics, 
Centre for Mathematical Sciences,\\ University of Cambridge, 
Wilberforce Road, Cambridge CB3 0WA, United Kingdom} 
\author{J. Menezes}  
\email{jmenezes@fc.up.pt} 
\affiliation{Centro de F\'{\i}sica do Porto, Rua do Campo Alegre 687, 4169-007 Porto, Portugal} 
\author{R. Menezes} 
\email{rms@fisica.ufpb.br} 
\affiliation{Departamento de F\'\i sica, Universidade Federal da Para\'\i ba,\\ 
 Caixa Postal 5008, 58051-970 Jo\~ao Pessoa, Para\'\i ba, Brazil} 
\author{J.C.R.E. Oliveira} 
\email{jeolivei@fc.up.pt} 
\affiliation{Centro de F\'{\i}sica do Porto, Rua do Campo Alegre 687, 4169-007 Porto, Portugal} 
\affiliation{Departamento de F\'{\i}sica da Faculdade de Ci\^encias 
da Universidade do Porto, Rua do Campo Alegre 687, 4169-007 Porto, Portugal} 

\date{13 December 2006}

\begin{abstract}
We describe the results of the largest and most accurate three-dimensional field theory simulations of domain wall networks with junctions. We consider a previously introduced class of models which, in the limit of large number $N$ of coupled scalar fields, approaches the so-called `ideal' model (in terms of its potential to lead to network frustration). We consider values of $N$ between $N=2$ and $N=20$. In all cases we find compelling evidence for a gradual approach to scaling, with the quantitative scaling parameters having only a mild dependence on $N$. These results strongly support our no-frustration conjecture.
\end{abstract}
\maketitle

\section{\label{sint}Introduction}

If our current understanding of particle physics and unification scenarios is correct, defect networks must necessarily have formed at phase transitions in the early universe \cite{KIBBLE}. The type of defect that forms and its specific properties depend on the particular details of each symmetry breaking, so a wide range of possibilities exist, with correspondingly different cosmological consequences \cite{VSH}.
 
Domain walls are usually pathological \cite{ZEL}, but it has been claimed \cite{SOLID} that if a domain wall network is frozen in comoving coordinates (or `frustrates', as is often colloquially put) then it can naturally explain the observational evidence that points to a recent acceleration of the universe. This happens because the equation of state of a domain wall gas is given by $w \equiv p/\rho = -2/3 + v^2$, $v$ being the root-mean squared (RMS) velocity of the domain walls. Note that we require that $w <-(1+\Omega_m^0/\Omega_{DE}^0) \lsim -1/2$ in order to accelerate the universe at the present time and consequently $v$ needs to be small. In practice the characteristic scale of the network, $L$, also needs to be tiny in order not to give rise to exceedingly large CMB fluctuations. The simplest domain wall models are known to reach a scaling regime (until they dominate the energy density of the universe), as first pointed out in \cite{PRESS} and recently studied in detail in \cite{AWALL,SIMS1,SIMS2}. Nevertheless, it was thought that more complicated models, notably those having junctions, would eventually frustrate.

In previous work \cite{IDEAL1,IDEAL2} we have studied the dynamics of domain wall networks with junctions, and investigated in detail energy, geometrical and topological constraints on the properties of domain wall networks. This led us to develop an `ideal' class of models which includes, in the large $N$ limit, what we called the `ideal' model (that is, the best candidate for frustration). However, a series of analytic and numerical arguments led us to a \textit{no frustration conjecture}: even though one can build (purely by hand) special lattices that would be locally stable against small perturbations \cite{BATTYE,CARTER,IDEAL2}, no such configurations are expected to ever emerge from any realistic cosmological phase transition. Our high-resolution numerical simulations of the ideal and other models showed clear evidence of a gradual approach to scaling, which was subsequently confirmed in \cite{BMFLAT}. 
 
Still, in our previous work \cite{IDEAL1,IDEAL2} we only considered numerical simulations in two spacetime dimensions. The present work is the third in this series of papers, and its goal is to eliminate this shortcoming. We report on the results of a series of massively parallel domain wall network numerical simulations of the `ideal' class of models in three spatial dimensions, which provide conclusive evidence for a gradual approach to scaling and hence strongly support our \textit{no frustration conjecture}. We start by discussing various dynamical issues relevant for the evolution of domain wall networks and briefly explain the `ideal' model. We then briefly describe our numerical implementation. Our code is based on the algorithm of Press, Ryden \& Spergel \cite{PRESS}, though with a number of crucial improvements that we will point out. Finally, we discuss our results and draw some conclusions. 


\section{\label{smod}The ideal domain wall model}

In \cite{AWALL} we introduced a phenomenological one-scale model for the 
evolution of domain wall networks that has been shown to provide a good 
approximation to the evolution of two key network parameters: the 
characteristic scale of the network, $L$, and the RMS velocity of the 
domain walls, $v$. The evolution equations are
\begin{equation}
\frac{dL}{dt}=HL+\frac{L}{\ell_d}v^2+c_wv\,,\ \  \frac{dv}{dt}=(1-v^2)\left(\frac{k_w}{L}-\frac{v}{\ell_d}\right)\,,
\label{rhoevoldw}
\end{equation}
where $H$ is the Hubble parameter, $c_w$ is the energy loss efficiency, 
$k_w$ is the curvature parameter and we have defined a damping length scale, $
1/\ell_d=3H+1/\ell_f$, which includes both the effects of Hubble damping and particle scattering.
The characteristic scale of the network is defined as $L=\sigma / \rho$
where $\rho$ is the average density in domain walls and $\sigma$ is the wall 
mass per unit area. Note that if domain 
walls are an important contribution to the dark energy then $\rho$ 
must be of the order of the critical density, $\rho_c$, at the present time.

If we ignore, for the moment, the effects of friction and the energy loss by 
the network (by making $\ell_f \to \infty$ and $c_w=0$) it possible to show 
that a linear scaling solution is possible for $a \propto t^{\alpha}$ with 
$\alpha > 1/4$. In the radiation era we obtain $L={\sqrt {4/3}} k_wt$ 
and $v=1/{\sqrt 3}$ while in the matter era we have 
$L={\sqrt {3/2}} k_wt$ and $v=1/{\sqrt 6}$.
We see that in both eras we have $L \sim k_wt$ and relatively large 
velocities. If we require CMB temperature fluctuations generated by domain walls on scales of the order of Hubble radius to be smaller than $10^{-5}$ then one would need $(LH)^{3/2} \lsim  10^{-5}$ or equivalently $L \lsim 1 \, {\rm Mpc}$ (our conservative estimate in \cite{IDEAL1}). However, CMB observations imply that the fluctuations generated by the domain walls have to be smaller than $10^{-5}$ down to much smaller scales ($\sim H^{-1}/100$). This means that current constraints on $L$ are expected to be roughly $2$ orders of magnitude 
stronger ($L \lsim 10 \, {\rm kpc}$) which implies a very small curvature 
parameter $k_w \lsim 10^{-6}$). This clearly shows that the simplest domain
wall scenario without junctions (with $k_w \sim 1$) is ruled out as a dark energy scenario. It is easy to show that allowing for a non-zero $c_w$ leads to a larger $L$ and consequently it does not help frustration \cite{AWALL,IDEAL1}. On the other hand, including friction also does not help much, due to the limited amount of energy with which domain walls can interact conserving energy and momentum \cite{IDEAL1}.

This failure of the simplest domain wall scenario led us to consider more complex scenarios with junctions and in \cite{IDEAL1} we investigated in detail energy, geometrical and topological considerations that severely constrain the properties of domain wall networks. In particular, in the context of 2D domain wall networks, we have shown using local energy considerations that two edge domains are always unstable and that three, four and five edge domains 
will be unstable if only Y-type junctions occur in a given model. We have also 
demonstrated that  increasing the average dimensionality of the junctions, 
$\langle d \rangle$, leads to a decrease of the average number of edges, 
$\langle x \rangle$, per domain (in particular if $\langle d \rangle > 6$ 
then $\langle x \rangle < 3$ and consequently no equilibrium configurations 
will ever form). Also, allowing for domain walls with different tensions 
contributes to increasing the instability since the walls with higher tension 
will tend to collapse thus increasing the dimensionality of the junctions 
which, in turn, will in general lead to the production of further unstable 
two edge domains.

Another important aspect, not yet discussed in previous papers, 
is related to the fact that the stability of a given domain depends on global considerations (those depending on the configuration of the surrounding domains) as well as local ones (those associated with the domain itself) and we expect the global ones to become more important as we increase the dimensionality of the junctions or consider specific domains with a large number of edges. In particular it is possible to show that a domain with three edges only survives the local stability analysis by very little in the case where there are only $X$ type junctions (the potential energy after the collapse would be at most about $10\%$ larger than before). In this case, we expect that, in general, non-local effects will make a three edge domain unstable. As a result in a model with only $X$ type junctions the only possible  stable configuration is the one in which all the domains have the same number of edges, $4$, which never occurs in the context of realistic domain wall network simulations. Note that we are assuming the junctions to be free throughout the paper. Otherwise their energy-momentum contribution could not be neglected, spoiling the dark energy properties associated with a static domain wall network.

The above energy, geometrical and topological considerations led us to propose a class of models with $N$ scalar fields and $N+1$ vacua with the property that all possible domain walls have equal tensions \cite{IDEAL2}. The `ideal' model is obtained in the limit $N \to \infty$. For large $N$ the collapse of a single domain will only very rarely lead to the fusion of two of the surrounding domains. This is clearly a very desirable feature from the point of view of frustration (see \cite{IDEAL1}). Also, by requiring all domain walls to have equal tensions we avoid another potential source of instability.

A specific realization of this class of models with $N$ scalar fields and 
a scalar field potential with $N+1$ vacua was given in \cite{IDEAL2}
\be
V \propto \sum_{j=1}^{N+1} r_j^2 \left(r_j^2 - r_0^2\right)^2\ {\rm with}\ 
r_j^2=\sum_{i=1}^N (\phi_i - p_{{i}_{j}})^2\,, \label{ideal}
\ee
where $p_{{i}_{j}}$ are the $N+1$ coordinates of the vacua of the potential. We have chosen $p_{{i}_{j}}$ to be the vertices of an ($N+1$)-dimensional regular polyhedra, and fixed the distance between the vacua to be equal to the parameter $r_0$. Given that in this model all possible domain walls have equal tensions, only Y-type junctions will form. Although it is possible to have only $X$ type junctions in a model with $4$ minima, this is in general not the case for a larger number of minima if all possible vacuum configurations exist in a given simulation. In particular this means that it is not possible to construct a generalization of the `ideal' model in which all junctions are of the $X$ type. 

In \cite{IDEAL2} we performed 2D domain wall network simulations of the `ideal' 
class of models with $N=4$ and $N=7$. These results are useful since they are much simpler (hence easier to understand), containing many important features which are also relevant in higher dimensions. However, realistic 3D domain wall network simulations are required in order to test our \textit{no frustration conjecture}. In the next section we shall briefly describe these simulations.


\section{\label{sfield}Massively parallel simulations}

We performed high-resolution field theory numerical simulations on the UK Computational Cosmology Consortium's COSMOS supercomputer using a modified version of the algorithm of Press, Ryden and Spergel \cite{PRESS}. These are described in detail in \cite{SIMS1,SIMS2}. The PRS algorithm \cite{PRESS} modifies the domain wall thickness in order to ensure a fixed comoving resolution. We measure the domain wall velocities using an algorithm analogous to that described in \cite{AWALL} which removes the radiated energy from the walls which otherwise would contaminate the estimate of the velocities. This is clearly an important advantage over previous velocity estimations (see for example \cite{PRESS}). We defined the domain wall as the region where $V(\phi) > \alpha \, V_{\rm max}$ ($V_{\rm max}$ being the maximum of the potential and $0 < \alpha < 1$) and we estimated the velocities as
\be
v_*^2 \equiv \langle v^2 \gamma^2 \rangle \sim \sum_{V(\phi_i) > \alpha V_{\rm max}}  \frac{{\dot \phi_i}^2}{2 V(\phi_i)}\, ,
\ee
where a dot represents a derivative with respect to conformal time and $\gamma=(1-v^2)^{1/2}$. The comoving characteristic scale of the network was estimated as $L_c \equiv L/a \sim \delta / f$ 
where $f$ is the volume fraction of the box with domain walls (that is with 
$V(\phi) > \alpha \, V_{\rm max}$ following our previous definition) and $\delta$ is the thickness of a static domain wall. We verified that, for $0.2 \lsim \alpha \lsim 0.6$, our results are almost independent of the threshold $\alpha$, as long as the domain wall is sufficiently resolved. We assume initial conditions where the scalar field at each point in the grid is associated with a randomly choosen minimum of the potential. 

The PRS algorithm was then parallelized with OpenMP directives, and optimized for the shared memory architecture of COSMOS. The results to be discussed below come from series of matter era simulations of $128^3$ and $256^3$ boxes, for all values of $N$ between 2 and 20, and with 10 different runs for each box size and number of fields. In addition, a single $512^3$ box was simulated for all $N$ between 2 and 20. A more extensive series of simulations studying, among other things, the evolution in other backgrounds, will be described in a subsequent publication. 

The required memory is approximately $DIM^3*N*4.5*8/1024^2$ MB, so for $512^3$ boxes up to 90 GB are required (for 20 fields). An output box binary file can also be produced at specified timesteps which can then be used to generate animations, an example of which is available at \url{http://www.damtp.cam.ac.uk/cosmos/viz/movies/evo2_25620_msmpeg.avi} As a benchmarking example, a $512^3$ simulation with 3 fields (requiring about 14.5 Gb of memory) takes about 14 seconds per step on 16 processors and only 5 seconds per step on 32 processors (as the memory ratio becomes favourable) without box output, and a complete run takes just 85 minutes. For larger runs the scalability is good if one keeps the memory smaller than 1 GB per processor. The largest simulation we have performed, a $512^3$ box with 20 scalar fields and box outputs at every timestep, took about 2 hours and 15 minutes on 128 processors.

\begin{figure}
\includegraphics[width=3in]{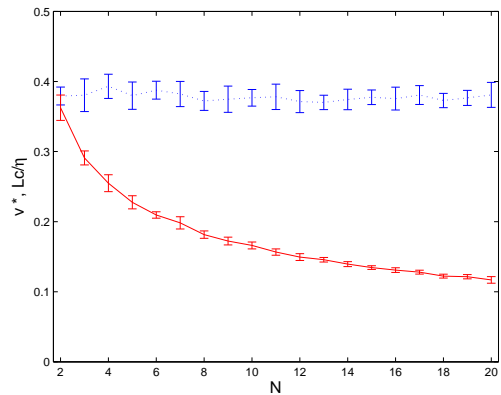}
\caption{\label{scalings} The asymptotic values of $v_*$ and $L_c/\eta$ for 
the ideal class of models with $N$ ranging from 
$2$ to $20$ (dotted and solid lines respectivelly). In Figs. \ref{scalings} 
and \ref{exponents} the error bars represent 
the standard deviation in an ensemble of 10 simulations.}
\end{figure}


\section{\label{sres}Results and discussion}

In Fig. \ref{scalings} we plot the asymptotic values of $v_*$ and $L_c/\eta$ 
(dotted and solid lines respectivelly) for the ideal class of models with 
$N$ ranging form 
$2$ to $20$ (we end our simulations when $\eta$ becomes equal to the box 
comoving size). The error bars represent the standard deviation in an 
ensemble of $10$ simulations. 
 As we increase $N$ the asymptotic value of $L_c/\eta$ 
decreases which is expected as we 
get closer to the ideal model. It is also significant that the differences between the 
successive $N$ results for $L_c/\eta$ become increasingly smaller for large $N$ 
which is a clear indication that the results obtained for $N=20$ are already close 
to the $N \to \infty$ results. On the other hand, we do not find any significant 
dependence the velocities, $v_*$, with $N$. 

\begin{figure}
\includegraphics[width=3in]{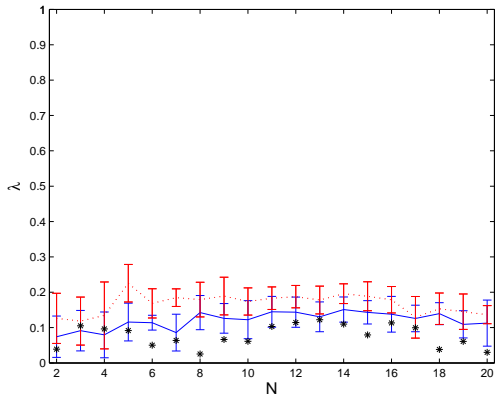}
\caption{\label{exponents}  The scaling exponents, $\lambda$, for all $N$'s between 2 and 20, for the $128^3$, $256^3$ boxes (dotted and solid lines 
respectivelly). The $*$'s represent the exponent measured for single $512^3$ boxes.
Note that frustration would correspond to $\lambda=1$.}
\end{figure}

In Fig. \ref{exponents} we plot the scaling exponents, $\lambda$, defined by 
$L_c/\eta \propto \eta^{-\lambda}$, for all values of $N$ between 
2 and 20, for the $128^3$, $256^3$ and boxes (dotted and solid lines 
respectivelly). The error bars represent the standard deviation in 
an ensemble of ten simulations and the $*$'s
represent the exponent measured for single $512^3$ boxes. The scaling 
exponents were computed using the results from the second quarter of the 
dynamical range of the simulations. We see that $\lambda$ is slightly 
greater than zero which indicates that there are small departures a the 
scaling solution. The fact that as we increase the box size, thus evolving 
the simulations for a longer dynamic range, $\lambda$ gets closer to zero 
is a clear indication that the networks are slowly aproaching 
a scaling solution. 
We have also performed 3D simulations in a model with only X-type junctions 
and $N=3$ ($4$ minima - see \cite{SIMS2} for specific realizations) and 
found no significant improvements with respect to the $N=2$ case ($3$ minima) 
with only Y-type junctions.

\section{\label{sdsc}Conclusions}

In this letter we presented the most compelling evidence to date that domain wall networks can not be the dark energy. Note that in order to be able to rule out the domain wall scenario for dark energy a very large and rich  class of models has to be analysed in detail. This lead us to develop a model best suited for frustration (the `ideal' model). We have shown that even this model fails to produce a frustrated domain wall network. Current observational constraints using cosmic microwave background and supernova data already strongly disfavour $w=-2/3$ as the equation of state of a single dark energy component \cite{WMAP3}. However, we should bear in mind that these results are dependent on strong priors. Even if we take them for granted and accept that domain walls alone cannot be the dark energy, they could still make a significant contribution. Our results, however, seem to exclude even that rather more contrived possibility. 

\section{acknowledgments}
This work was done in the context of the ESF COSLAB network and funded by FCT (Portugal), in the framework of the POCI2010 program, supported by FEDER. Specific funding came from grant POCI/CTE-AST/60808/2004, SFRH/BSAB/603/2006 (P.A.) and from the Ph.D. grant SFRH/BD/4568/2001 (J.O.). C.M. is grateful to the Galileo Institute for Theoretical Physics for hospitality, and to the INFN for partial support during the completion of this work. J.M. and R.M. are supported by the Brazilian government (through CAPES-BRASIL), specifically through grants BEX-1970/02-0 and BEX-1090/05-4.

The numerical simulations were performed on COSMOS, the Altix3700 owned by the UK Computational Cosmology Consortium, supported by SGI, Intel, HEFCE and PPARC. The support of the COSMOS parallel programmer, Victor Travieso, with the code optimization and data visualization is greatly appreciated.


\bibliography{prl}

\end{document}